\documentclass[12pt]{iopart}
\usepackage{iopams}
\usepackage{graphicx}
\usepackage[colorlinks=true]{hyperref}

\begin{document}

\title{Non-spreading matter-wave packets in a ring}

\author{Jieli Qin}

\address{School of Physics and Electronic Engineering, Guangzhou University,
230 Wai Huan Xi Road, Guangzhou Higher Education Mega Center, Guangzhou
510006, China}
\eads{\mailto{qinjieli@126.com}, \mailto{104531@gzhu.edu.cn}}

\vspace{10pt}
\begin{indented}
\item[] \today
\end{indented}

\begin{abstract}
Non-spreading wave packets and matter-wave packets in ring traps both
have attracted great research interests due to their miraculous physical
properties and tempting applications for quite a long time. Here,
we proved that there exists only one set of non-spreading matter-wave
packets in a free ring, and this set of wave packets have been found
analytically. These non-spreading matter-wave packets can be realized
in a toroidal trapped Bose-Einstein condensate system with the help
of Feshbach resonance to eliminate contact interaction between atoms.
Since experimentally residual interaction noise will always exist,
its effect on the stability of these non-spreading wave packets is also
examined. Qualitatively, under weak residual interaction noise, these
non-spreading wave packets can preserve their shape for quite a long
time, while a stronger interaction noise will induce shape breathing
of the wave packets. Shape-keeping abilities of these wave packets
are further studied quantitatively. We found that this set of wave
packets have the same shape-keeping ability against interaction noise.
And, the shape-keeping ability is linearly related to the interaction
noise strength.
\end{abstract}

\vspace{2pc}

\submitto{\PS}

%\maketitle

\section{Introduction}

Matter-waves with substantial intensity are crucial in applications
such as atom lithography \cite{Gangat2005,Fouda2016,Dwyer2005},
ultra-sensitive magnetometry \cite{Vengalattore2007,Muessel2014},
matter-wave interferometry \cite{Cronin2009,Lee2012,Muntinga2013}.
When a matter-wave packet propagates in free space, it will spread
due to dispersion, and its intensity will decrease \cite{Robins2013,Bolpasi2014}.
Thus, non-spreading matter-wave packets attract great research interests.
It is well known that self-focusing nonlinearity can be introduced to
overcome the dispersion spreading, and form bright matter-wave
solitons \cite{Khaykovich2002,Strecker2002,Strecker2003,Abdullaev2008} which
are non-spreading wave packets. Soliton-based matter-wave interferometers
have been elaborated in many theoretical
works \cite{Martin2012,Polo2013,Gertjerenken2013,Helm2014,Sakaguchi2016},
and also experimentally implemented with an increase of interference fringe
visibility been observed \cite{McDonald2014,McDonald2017}. However,
nonlinearity may also do harm to the performance of interferometers,
it would induce a phase diffusion, which will reduce the coherent
time \cite{Fattori2008,Gustavsson2008}. And recently it is also reported
that the best sensitivity of a matter-wave interferometer is reached
in the linear regime \cite{Haine2016,Haine2018}. (We note that the
relation between nonlinearity and performance of an interferometer
is quite subtle. Besides the above-mentioned effects, nonlinearity
also gives rise to non-classical correlations and squeezed
states \cite{Jo2007,Berrada2013}. Taking advantage of these properties,
standard quantum limit surpassed interferometers can be
realized \cite{Gross2010,Lucke2011}.) So, achieving linear non-spreading
matter-wave packets can be a charming research subject.

In 1979, Berry and Balazs showed that the free particle linear Schr\"{o}dinger
equation has a nontrivial Airy function solution which holds its
shape during an accelerating propagation \cite{Berry1979}. After
this pioneering work, such linear non-spreading waves were extensively
studied (for a review see reference \cite{Hu2012}), and first experimentally
demonstrated in optical system in 2007 \cite{Siviloglou2007}.
Soon after, linear non-spreading matter-waves were also realized in
electron beams \cite{Voloch-Bloch2013}. Atomic Bose-Einstein condensate
(BEC), because of its macroscopic quantum properties and highly controllable
feature, is an ideal system for exploring matter-wave optical phenomena.
Recently, generation of similar non-spreading BEC wave packets by
amplitude or phase imprinting techniques \cite{Efremidis2013}, and
time dependent harmonic traps \cite{Yuce2015} have also been proposed.

Due to potential applications in realizing matter-wave Sagnac interferometer
\cite{Bouyer2014,Barrett2014,Dayon2010,Helm2015,Stevenson2015}, atomic analogy
of SQUID circuits \cite{Ryu2013,Wang2015,Mathey2016}, persistent
current \cite{Kumar2016,Beattie2013,Ryu2007,Brand2001}
and quantum time crystal \cite{Li2012,Ohberg2018,Sacha2018}, non-spreading
matter-wave packets in a ring also deserve considerable research interests.
Now the point is that, we are inquisitive about whether non-spreading
wave packets can exist in a free ring. In this article, first we analytically
found a set of non-spreading wave packets in a free ring, at the same
time we also proved that they are the only set. Then the realization
of these non-spreading wave packets in atomic BEC system with the help
of Feshbach resonance \cite{Chin2010,Timmermans1999} technique to
eliminate inter-atom interaction is discussed. At last, although in
principle inter-atom interaction can be totally eliminated by
Feshbach resonance technique, practically it is impossible to operate the Feshbach
magnetic field with infinite precision, and there will
always exist some residual interactions \cite{Fattori2008,Gustavsson2008,Pollack2009},
therefor the stability of these non-spreading wave
packets against residual interaction noise is also studied.

The paper is organized as follows. In section
\ref{sec:Non-spreading}, formulas describing non-spreading wave packets
in a ring are derived, and the only set of non-spreading matter-wave
packets are found analytically. In section \ref{sec:Realization},
the experimental realization of the non-spreading wave packets in
BEC systems are discussed briefly. In section \ref{sec:Stability},
the stability of the wave packets against residual interaction noise
is studied numerically. At last, the work is summarized in section
\ref{sec:Conclusion}.

\section{Non-spreading matter-wave packets in a ring\label{sec:Non-spreading}}

To generally discuss the problem of non-spreading wave packets in
a ring, we begin with the following dimensionless Schr\"{o}dinger
equation
\begin{equation}
i\frac{\partial}{\partial t}\psi\left(\theta,t\right)
  =-\frac{1}{2}\frac{\partial^{2}}{\partial\theta^{2}}\psi\left(\theta,t\right),
  \label{eq:Schrodinger}
\end{equation}
where $\theta$ is the azimuthal angle which takes value in range
of $-\pi$ to $\pi$. Here we are interested in the case of a free
ring, therefore in the equation there only exists the kinetic energy
term, while external potential term is not included. The continuous
of wave function and its first order derivative require
\begin{equation}
\psi\left(\theta_{0},t\right)=\psi\left(\theta_{0}+2n\pi,t\right),\label{eq:boundary0}
\end{equation}
and
\begin{equation}
\psi'\left(\theta_{0},t\right)=\psi'\left(\theta_{0}+2n\pi,t\right),\label{eq:boundary1}
\end{equation}
where $\theta_{0}$ is an arbitrary amizuthal angle, $n=0,\pm1,\pm2,\cdots$
is an integer number, and $\psi'\left(\theta_{0},t\right)=
\left.\frac{\partial\psi\left(\theta,t\right)}{\partial\theta}\right|_{\theta=\theta_{0}}$
is the first order derivative of $\psi\left(\theta,t\right)$ at point
$\theta=\theta_{0}$.

To find a shape preserving wave packet, we rewrite the wave function
$\psi\left(\theta,t\right)$ as follows \cite{Madelung1927,Lin2008},
\begin{equation}
\psi\left(\theta,t\right)=A\left(\theta,t\right)e^{iS\left(\theta,t\right)},\label{eq:madelungForm}
\end{equation}
 where $A$ and $S$ are two real number functions, representing the
probability density and phase distribution respectively. Under such
a form, if function $A\left(\theta,t\right)$ can be expressed as
$A\left(\theta,t\right)=A\left(\phi\right)$ with
\begin{equation}
\phi=\theta-f\left(t\right),\label{eq:trajectory}
\end{equation}
the wave packet will preserve its shape during the propagation. Here
$\phi$ can be regarded as the trajectory function of the wave packet,
with $f\left(t\right)$ being a real number function.

Inserting the Madelung transformed wave function (\ref{eq:madelungForm})
into Schr\"{o}dinger equation (\ref{eq:Schrodinger}), and splitting
real and imaginary parts of the equation, we get the following two
equations
\begin{equation}
-\frac{\partial A}{\partial\phi}\frac{df}{dt}=-\frac{1}{2}\left[2\frac{\partial A}{\partial\phi}\frac{\partial S}{\partial\theta}+A\frac{\partial^{2}S}{\partial\theta^{2}}\right],\label{eq:ImEq}
\end{equation}
and
\begin{equation}
-A\frac{\partial S}{\partial t}=-\frac{1}{2}\left[\frac{\partial^{2}A}{\partial\phi^{2}}-A\left(\frac{\partial S}{\partial\theta}\right)^{2}\right].\label{eq:ReEq}
\end{equation}
The first equation (\ref{eq:ImEq}) can be rearranged into the following
form
\[
\frac{\partial}{\partial\theta}\left(\frac{df}{dt}A^{2}-\frac{\partial S}{\partial\theta}A^{2}\right)=0,
\]
 which means that $\frac{df}{dt}A^{2}-\frac{\partial S}{\partial\theta}A^{2}$
is independent of azimuthal angle $\theta$. So we can write it as
\begin{equation}
\left[\frac{df\left(t\right)}{dt}-\frac{\partial S\left(\theta,t\right)}{\partial\theta}\right]
A^{2}\left(\phi\right)=c\left(t\right),\label{eq:condition1}
\end{equation}
with $c\left(t\right)$ being an arbitrary function depending on time
variable $t$ only.

First, it can be obviously seen that equation (\ref{eq:condition1})
admits a trivial solution
\[
A\left(\phi\right)=cons,\quad S\left(\theta,t\right)\propto\theta.
\]
Equation (\ref{eq:ReEq}) and wave function continuous condition (\ref{eq:boundary0},
\ref{eq:boundary1}) also considered, corresponding wave function
reads
\begin{equation}
\psi\left(\theta,t\right)=Ce^{il\theta+il^{2}t/2},\label{eq:trivialSol}
\end{equation}
which is a trivial solution with uniformly distributed density in
the ring at all time.

If $A\left(\phi\right)$ is not a constant, by adapting its background
value, there always exists an angle $\phi_{0}$ making $A\left(\phi_{0}\right)=0$.
And note that the right-hand side of equation (\ref{eq:condition1})
is only a function of $t$, this is to say for all $\phi$ the following
equation need to be satisfied
\[
\left[\frac{df\left(t\right)}{dt}-\frac{\partial S\left(\theta,t\right)}{\partial\theta}\right]
A^{2}\left(\phi\right)=c\left(t\right)=0.
\]
From this equation, we get the following relationship between $f\left(t\right)$
and $S\left(\theta,t\right)$
\begin{equation}
\frac{df\left(t\right)}{dt}=\frac{\partial S\left(\theta,t\right)}{\partial\theta}
=\sum_{i=0}^{\infty}c_{i}t^{i},\label{eq:series}
\end{equation}
here since $\frac{df\left(t\right)}{dt}$ is a function only depending
on time variable $t$, we expand it in Taylor series with $c_{j}$
being the $j$-th order coefficient. In the very following content,
depending on the highest order of this series we discuss the problem
in three distinct cases:
\begin{enumerate}
\item \label{item1} If series (\ref{eq:series}) only has the constant term, i.e.,
\[
\frac{df\left(t\right)}{dt}=\frac{\partial S\left(\theta,t\right)}{\partial\theta}=c_{0},
\]
then it is straightforward to get
\[
f\left(t\right)=c_{0}t,
\]
\[
S\left(t\right)=c_{0}\theta+g\left(t\right),
\]
with $g\left(t\right)$ being an arbitrary function. Substituting
them into equation (\ref{eq:ReEq}), we have
\[
-A\left(\phi\right)\frac{\partial S\left(\theta,t\right)}
{\partial t}=-\frac{1}{2}\left[\frac{\partial^{2}A\left(\phi\right)}
{\partial\phi^{2}}-A\left(\phi\right)c_{0}^{2}\right].
\]
Because the right-hand side of this equation does not explicitly depend
on time variable $t$, the equality requires
$\frac{\partial S\left(\theta,t\right)}{\partial t}=\frac{dg\left(t\right)}{dt}$
is also explicitly time independent. That is to say, function $g\left(t\right)$
must take the following linear form
\[
g\left(t\right)=G_{1}t+G_{0},
\]
where $G_{0}$ and $G_{1}$ are two free constant parameters. At last,
equation (\ref{eq:ReEq}) becomes
\[
\frac{\partial^{2}A\left(\phi\right)}{\partial\phi^{2}}+\left(c_{0}^{2}-2G_{1}\right)A\left(\phi\right)=0,
\]
from which the shape of a non-spreading wave packet can be determined
\[
A\left(\phi\right)=C\cos\left(m\phi\right).
\]
Here $m=\sqrt{\left|c_{0}^{2}-2G_{1}\right|}$ is an integer number
which is restricted by the boundary condition, $C$ is a normalization
constant. At the same time, boundary conditions also require that
$c_{0}$ be an integer number $c_{0}=l$. At last, putting the above
expressions of $f,g,A,S$ all together, a full non-spreading solution
can be constructed as
\begin{eqnarray}
\psi\left(\theta,t\right)= & C\cos\left(m\theta-lt\right)\nonumber \\
 & \exp\left[il\theta+i\frac{\left(m^{2}+l^{2}\right)}{2}t\right].\label{eq:solution}
\end{eqnarray}
The physical meanings of $m$ and $l$ are obvious. $2m$ is
number of nodes of the wave function, and $l/m$ describes the moving
speed of the wave packet along the ring. We also note that the previous
trivial solution (\ref{eq:trivialSol}) is just the $m=0$ specification
of solution (\ref{eq:solution}).
\item \label{item2} If series (\ref{eq:series}) has terms up to the first order, i.e.,
\[
\frac{df\left(t\right)}{dt}=\frac{\partial S\left(\theta,t\right)}{\partial\theta}=c_{0}+c_{1}t,
\]
function $f\left(t\right)$ and $S\left(\theta,t\right)$ are integrated
to be
\[
f\left(t\right)=c_{0}t+\frac{1}{2}c_{1}t^{2},
\]
\[
S\left(\theta,t\right)=c_{0}\theta+c_{1}\theta t+g\left(t\right).
\]
Then equation (\ref{eq:ReEq}) becomes
\begin{eqnarray*}
 & \frac{1}{2}\frac{\partial^{2}A\left(\phi\right)}{\partial\phi^{2}}
 -c_{1}\phi A\left(\phi\right)-\frac{1}{2}c_{0}^{2}A\left(\phi\right)\\
= & A\left(\phi\right)\left[g'\left(t\right)+2c_{0}c_{1}t+c_{1}^{2}t^{2}\right].
\end{eqnarray*}
As in case (\ref{item1}), the left-hand side of this equation also does not explicitly
depend on variable $t$. And in this case, equality between the left
and right-hand sides can be met by the following function of $g\left(t\right)$,
\[
g\left(t\right)=-\frac{1}{3}c_{1}^{2}t^{3}+c_{0}c_{1}t^{2}+G_{1}t+G_{0}.
\]
At last, the equation governing function $A\left(\phi\right)$ reads
\[
\frac{1}{2}\frac{\partial^{2}A\left(\phi\right)}{\partial\phi^{2}}
-c_{1}\phi A\left(\phi\right)-\left(\frac{1}{2}c_{0}^{2}+G_{1}\right)A\left(\phi\right)=0,
\]
which promises an Airy function solution. But an Airy function cannot
 fulfill the periodical boundary condition (\ref{eq:boundary0})
and (\ref{eq:boundary1}), therefore a non-spreading wave packet cannot
 exist in this situation.
\item If the series have terms equal or higher than second order, i.e.,
\[
\frac{df\left(t\right)}{dt}=\frac{\partial S\left(\theta,t\right)}{\partial\theta}=c_{0}+c_{1}t+c_{2}t^{2}+\cdots,
\]
we have
\[
f\left(t\right)=c_{0}t+\frac{1}{2}c_{1}t^{2}+\frac{1}{3}c_{2}t^{3}+\cdots,
\]
\[
S\left(\theta,t\right)=\theta\left(c_{0}+c_{1}t+c_{2}t^{2}+\cdots\right)+g\left(t\right).
\]
Similar as in case (\ref{item1}) and (\ref{item2}), we rewrite equation (\ref{eq:ReEq}) as
follows
\[
\frac{1}{2}\frac{\partial^{2}A}{\partial\phi^{2}}-c_{1}\phi A-\frac{1}{2}c_{0}^{2}A=A\left[\begin{array}{c}
\left(c_{0}c_{1}+2c_{2}\phi\right)t\\
+\left(c_{1}^{2}+2c_{0}c_{2}\right)t^{2}\\
+\cdots+g'\left(t\right)
\end{array}\right]
\]
The left-hand side of this equation is still $t$ independent. But
because of the existence of new term $2c_{2}\phi t$ (and some other
higher order ones), there does not exists any function $g\left(t\right)$
which can make the right-hand side also being time independent. Thus,
this equation has no solution. There does not exist non-spreading
wave packet solutions in such a case.
\end{enumerate}
Now, in a brief summary, combing all the three above cases, we can
conclude that the only set of non-spreading wave packets in a ring
is given by equation (\ref{eq:solution}). This set of non-spreading
wave packets travel along the ring with a constant velocity. Moreover,
owing to the quantum feature, these non-spreading wave packets can
only propagate in some quantized fractional velocities. And unlike
in free space, self-accelerating non-spreading wave packets with Airy
function like form cannot exist in a ring configuration.

\section{Realization in a toroidal trapped BEC system\label{sec:Realization}}

\begin{figure}
\begin{centering}
\includegraphics{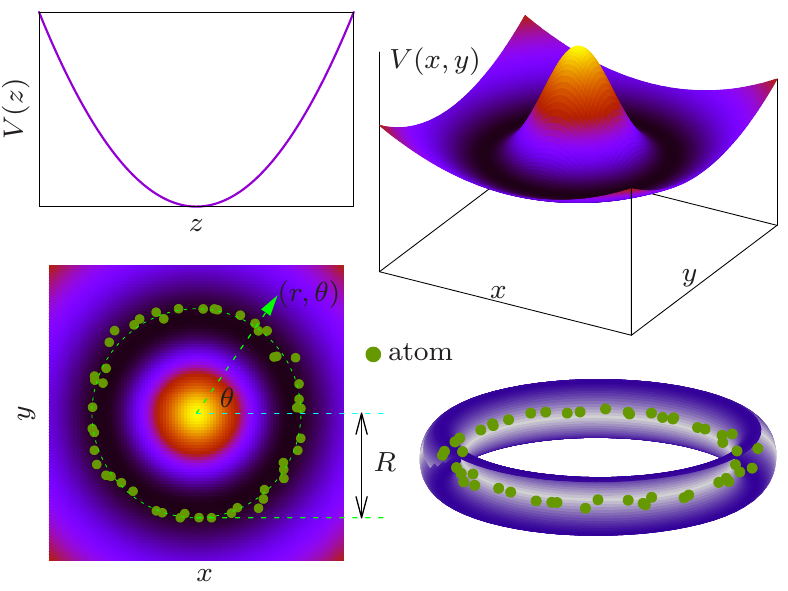}
\par\end{centering}
\caption{Diagram of toroidal trapped BEC. By applying a harmonic trap
$V\left(z\right)=m\omega_{z}^{2}z^{2}/2$
along $z$-direction, a two-dimensional harmonic trap and a Gaussian
barrier $V\left(x,y\right)=V\left(r,\theta\right)
=m\omega^{2}r^{2}/2+V_{0}\exp\left(-2r^{2}/w_{0}^{2}\right)$
in $x$-$y$ plane, BEC can be trapped in a ring with radius $R$
determined by formula $R^{2}=
w_{0}^{2}/2\ln\left[4V_{0}/\left(m\omega^{2}w_{0}^{2}\right)\right]$.
If the trap along $z$ and $r$ direction is sufficiently strong,
the system can be reduced to one-dimension, leaving only the space
variable $\theta$ to take effect. \label{fig:Sketch}}
\end{figure}

The non-spreading matter-wave packets found in section \ref{sec:Non-spreading},
could be experimentally realized in a quasi-one-dimensional toroidal
trapped dilute atomic BEC system, see figure \ref{fig:Sketch}. In
a BEC system, because of collision between atoms, besides the kinetic
energy and external trap potential, there will also exists a nonlinear
term, i.e., the system is described by the following nonlinear Schr\"{o}dinger
equation (Gross-Pitaevskii equation)
\begin{equation}
i\hbar\frac{\partial}{\partial t}\psi=
-\frac{\hbar^{2}}{2m}\nabla^{2}\psi
+V\left(r,\theta,z\right)\psi+g\left|\psi\right|^{2}\psi,\label{eq:GP}
\end{equation}
where $\hbar$ is the reduced Planck constant, $m$ is the mass of
atom, $g=4\pi\hbar^{2}a_{s}/m$ is the interaction strength with $a_{s}$
being the s-wave scattering length, and
\begin{equation}
V\left(r,\theta,z\right)=\frac{1}{2}m\omega_{z}^{2}z^{2}
+\frac{1}{2}m\omega^{2}r^{2}+V_{0}\exp\left[-\frac{2r^{2}}{w_{0}^{2}}\right],
\label{eq:Potential}
\end{equation}
is the toroidal trapping potential which is experimentally realized
by a magnetic trap and a plug laser beam with Gaussian shape \cite{Ryu2007} (there are also some
other schemes to realize a toroidal trap, such as the ones
in ref \cite{Wright2000,Arnold2004,Crookston2005,Heathcote2008}, \emph{et al}.)
having its minimum at $z=0$ and $r=R$, with
\begin{equation}
R^{2}=\frac{w_{0}^{2}}{2}\ln\left(\frac{4V_{0}}{m\omega^{2}w_{0}^{2}}\right).\label{eq:RingR}
\end{equation}
Expanding the trapping potential around $r=R$ into Taylor series,
and neglecting the third and higher orders of $r-R$, the potential
approximately reads
\begin{equation}
V\left(r,\theta,z\right)=\frac{1}{2}m\omega_{z}^{2}z^{2}
+\frac{1}{2}m\omega_{r}^{2}\left(r-R\right)^{2},\label{eq:AppRingPoential}
\end{equation}
with $\omega_{r}=2\omega R/w_{0}$ being the effective radial trapping
frequency.

For sufficiently strong confinement in axial and radial direction,
the wave function can be assumed to have variables separated form
$\psi=\rho_{0}\left(r\right)\phi_{0}\left(z\right)\psi\left(\theta,t\right)$
with $\phi_{0}=\left[m\omega_{z}/\left(\pi\hbar\right)\right]^{1/4}
\exp\left[-\frac{m\omega_{z}z^{2}}{2\hbar}\right]$,
$\rho_{0}=\left[m\omega_{r}/\left(\pi\hbar r^{2}\right)\right]^{1/4}
\exp\left[-\frac{m\omega_{r}\left(r-R\right)^{2}}{2\hbar}\right]$
being ground state wave function of the axial and radial traps. Inserting
it into equation (\ref{eq:GP}), the system can be reduced to one-dimension.
 The azimuthal wave function $\psi\left(\theta,t\right)$
obeys equation
\begin{equation}
i\hbar\frac{\partial}{\partial t}\psi=-\frac{\hbar^{2}}{2mR^{2}}
\frac{\partial^{2}}{\partial\theta^{2}}\psi+\tilde{g}\left|\psi\right|^{2}\psi.
\label{eq:RingGP}
\end{equation}
Here $\tilde{g}=m\sqrt{\omega_{z}\omega_{r}}g/\left(2\pi\hbar R\right)$
is the quasi-one-dimensional effective contact interaction strength.
Taking transformation $t\rightarrow mR^{2}t/\hbar$, equation (\ref{eq:RingGP})
can be rescaled to the following dimensionless form
\begin{equation}
i\frac{\partial}{\partial t}\psi=-\frac{1}{2}\frac{\partial^{2}}{\partial\theta^{2}}
\psi+g_{s}\left|\psi\right|^{2}\psi,\label{eq:DimensioinlessRingGP}
\end{equation}
with $g_{s}=\tilde{g}mR^{2}/\hbar^{2}=2mRNa_{s}\sqrt{\omega_{r}\omega_{z}}/\hbar$.
Considering, for instance, BEC of $^{39}K$ atoms, with atoms number $N=10^3$,
s-wave scattering length $a_s=33a_0$ ($a_0$ is Bohr radius), trapped in a toroidal
trap with parameters $R=10\mu m, \omega_r=\omega_z=1000Hz$, the interaction
strength is $g_s \approx 22$. Thus, usually the interaction will play a
significant role. Fortunately, it can be eliminated by Feshbach resonance
technique. According to Feshbach resonance theory \cite{Chin2010,Timmermans1999},
s-wave scattering length $a_{s}$ can be tuned by applying a magnetic
field
\begin{equation}
a_{s}\left(B\right)=a_{s,\infty}\frac{B-B_{0}}{B-B_{r}},\label{eq:Feshbach}
\end{equation}
where $a_{s,\infty}$ is asymptotic s-wave scattering length in the
case of far from resonance, $B$ is the magnetic induction of the
applied magnetic field, $B_{r}$ is the resonant magnetic induction,
and $B_{0}$ is the value of magnetic induction for a vanishing s-wave
scattering. This is to say inter-atom interaction of BEC can be totally
eliminated when a magnetic field whose induction equals $B_{0}$ is
applied. And we get the same equation as (\ref{eq:Schrodinger}) in
section \ref{sec:Non-spreading}. So, we propose the non-spreading
wave packets can be realized in a toroidal trapped BEC system.

\section{Stability against residual interaction noise \label{sec:Stability}}

According to Feshbach resonance theory \cite{Chin2010,Timmermans1999},
in principle strength of the
inter-atom interaction can be tuned to 0 by a magnetic field $B_{0}$.
However, practically noise of magnetic field around $B_{0}$ is unavoidable,
thus a small residual interaction noise will always be left
\begin{equation}
g_{s}=g_{\xi}\xi\left(t\right).\label{eq:g_noise}
\end{equation}
Here $g_{\xi}$ is the strength of the residual interaction noise,
$\xi\left(t\right)$ is a random function. White noise assumed, $\xi\left(t\right)$
is uniformly distributed in range of $\left[-1,1\right]$. According to
experiment \cite{Fattori2008}, the interaction can be reduced by
a factor of $10^3$ with a 100 mG magnetic field stability, so a typical
strength of the residual interaction noise is $g_{\xi} \approx 0.022$
(recall that without Feshbach resonance the interaction strength is
estimated to be $g_s=22$ in the previous section). If the magnetic
field is operated more precisely (in experiment \cite{Gustavsson2008},
the magnetic field can be controlled to about 1 mG),
an even smaller value of $g_{\xi}$ can be obtained. Thus,
here we typically choose $g_{\xi}$ in order of $10^{-2}$,
and to study the limit behaviors, $g_{\xi}=0.002$
and $g_{\xi}=0.5$ are also examined.

It is then necessary to study the stability of those non-spreading matter-wave packets
found in section \ref{sec:Non-spreading} against such noises. Here,
this is done by numerically solving equation (\ref{eq:DimensioinlessRingGP})
with an operator splitting method. The nonlinear part is directly
integrated (since this is a stochastic integral, Ito stochastic integral
formula \cite{Gardiner1985} is used) in $x$-space, while the second
order differential term is handled in momentum space by means of
fast Fourier transformation.

\begin{figure*}
\begin{centering}
\includegraphics{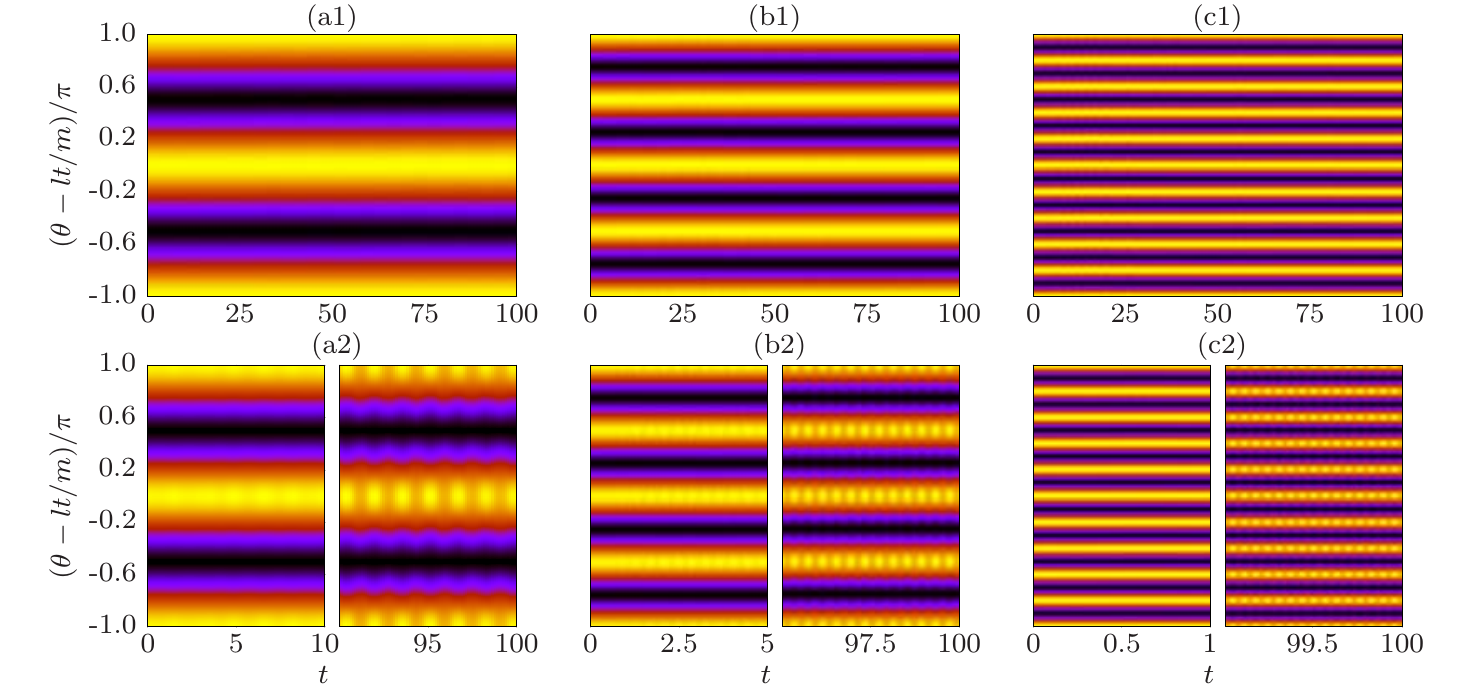}
\par\end{centering}
\caption{Evolution of some non-spreading wave packets under the influence of
residual interaction noise. Under weak interaction noise with strength
$g_{\xi}=0.002$ (figures a1-c1) the wave packets keep its shape during
the evolution, while a stronger interaction noise with strength $g_{\xi}=0.05$
(figures a2-c2) will induce shape oscillation of the wave packets.
Wave packet parameters are: $m=1,l=1$ for figures (a1, a2); $m=2,l=3$
for figures (b1, b2) and $m=5,l=5$ for figures (c1, c2).\label{fig:Examples}}
\end{figure*}

In figure \ref{fig:Examples}, we plot the evolution of some non-spreading
wave packets (with $m,l=1,1;2,3$ and $5;5$) under the influence
of residual interaction noise with both a very weak strength $g_{\xi}=0.002$
and a little stronger one $g_{\xi}=0.05$. (Here, we mean stronger compared
to $g_{\xi}=0.002$. But, it is still a weak interaction in fact.
The ``kinetic'' energy of a non-spreading wave packet is
$E_k=-\frac{1}{2}\int_0^{2\pi}\psi_{ml}^{*}\frac{\partial^2}{\partial \theta^2}
\psi_{ml}d\theta=l^2+m^2$, while the interaction energy is
$E_i=\frac{g_{\xi}}{2}\int_{0}^{2\pi}|\psi_{ml}|^4 d\theta
=\frac{3g_{\xi}}{8\pi}$. Thus,
for $g_{\xi}=0.05$, $E_i \ll E_k$ still holds.) In the figure,
to conveniently compare the shapes of a wave packet at different
times during the evolution, the vertical coordinate is
set to $\phi=\theta - lt/m$, i.e., the
wave packet is shifted to its initial location. From this figure,
we see that for very weak noise, the non-spreading wave packet can
keep its shape for quite a long time, while a stronger noise will
induce a periodical breathing of the wave packet shape after long time.
For both cases, the periodical travel of wave-packet in the ring is
not affected by the interaction noise.

\begin{figure}
\begin{centering}
\includegraphics{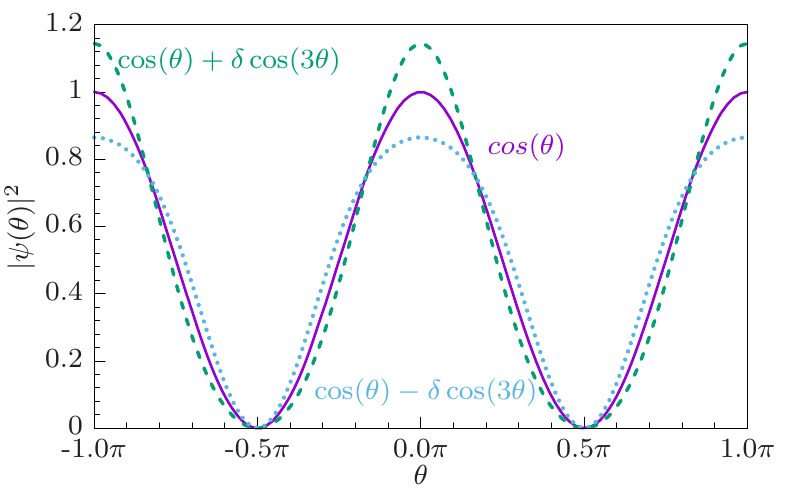}
\par\end{centering}
\caption{Schematic formation of a breathing mode excitation on the non-spreading
wave packet with $m=1$. The superposition of main wave function
$\psi=\cos\left(\theta\right)$
and small excitation wave function
$\Delta\psi=\delta\cos\left(\theta\right)\exp\left(i\varphi\right)$
forms a breathing mode. When $\psi$ and $\delta\psi$ have the same
phase, the wave packet is suppressed; while their phases are opposite,
the wave packet is broadened. \label{fig:Breathing}}
\end{figure}

\begin{figure}
\begin{centering}
\includegraphics{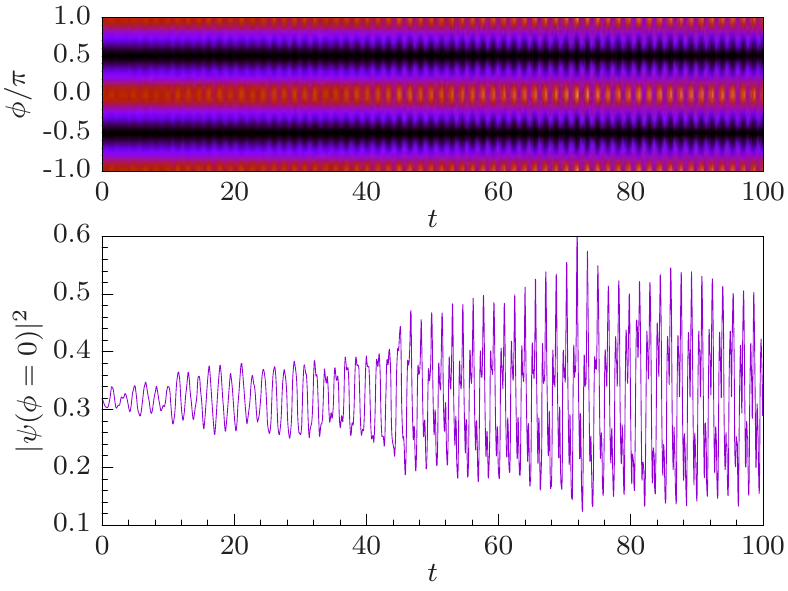}
\par\end{centering}
\caption{High order oscillating modes excited by interaction noise
with a considerable large strength $g_{\xi}=0.5$. The top panel is
a heat map plot of $|\psi(\phi,t)|^2$ with $\phi=\theta-lt/m$. The
bottom panel is a plot of $|\psi(\phi=0,t)|^2$. The wave packet
parameters are $m=1,l=1$. \label{fig:LargeNoise}}
\end{figure}

The numerically found shape breathing periods for different wave packets
are listed in table \ref{tab:ShapeBreathingPeriod}. From the table,
it is natural to conclude that breathing period is totally determined
by the wave packet parameters $m$, and has nothing to do with $l$
and $g_{\xi}$, and its value is approximately $T\approx\pi/\left(2m^{2}\right)$.
This oscillating period can be easily understood by considering the
formation mechanism of a breathing mode. For wave packet with quantum
number $m$, the interaction noise will induce a transition to state
with quantum number $3m$, thus excites a breathing mode. Taking $m=1$
for an example, in figure \ref{fig:Breathing}, we show the formation
of such a breathing mode by adding wave packet $\psi_{1}=\cos\left(\theta\right)$
and a small perturbation
$\Delta\psi_{1}=\delta\cos\left(3\theta\right)\exp\left[i\varphi\right]$.
When $\psi_{1}$ and $\Delta\psi_{1}$ are in phase ($\varphi=0$),
the excitation will have a suppressing effect on width of the wave
packet, while $\psi_{1}$ and $\Delta\psi_{1}$ have a phase difference
of $\pi$ ($\varphi=\pi$), the excitation will spread the wave packet.
Thus, for wave packet with quantum number $m$, the breathing period
will be
\begin{equation}
T=\frac{2\pi}{\left(3m-m\right)^{2}}=\frac{\pi}{2m^{2}}.\label{eq:Period}
\end{equation}
And if there exists a considerable large interaction noise, higher order oscillating
modes will be excited as shown in figure \ref{fig:LargeNoise} where the evolution
of non-spreading wave packet with $l=1,m=1$ under residual interaction noise with
strength $g_{\xi}=0.5$ is plotted. The heat map is a plot of $|\psi(\phi,t)|^2$ with
$\phi=\theta - lt/m$, and the line in the bottom panel is a plot of $|\psi(\phi=0,t)|$.
From both plots, higher frequency oscillations can be easily observed after about $t=40$.

\begin{table*}
\begin{centering}
\begin{tabular}{cc|c|cc}
\hline
\multicolumn{2}{c|}{Wave packet parameters} & Noise Strength & \multicolumn{2}{c}{Breathing Period}\\
\hline
$m$ & $l$ & $g_{\xi}$ & \multicolumn{2}{c}{$T$}\\
\hline
\hline
 &  & 0.01 & 1.5674 & \\
1 & 1 & 0.02 & 1.5678 & $\approx\pi/2$\\
 &  & 0.05 & 1.5678 & \\
\hline
 & 2 & 0.02 & 0.3934 & \\
2 & 3 & 0.05 & 0.3930 & $\approx\pi/8$\\
 & 4 & 0.05 & 0.3932 & \\
\hline
3 & 2 & 0.05 & 0.1742 & $\approx\pi/18$\\
\hline
5 & 5 & 0.05 & 0.0628 & $\approx\pi/50$\\
\hline
\end{tabular}
\par\end{centering}
\caption{Shape breathing period of some non-spreading wave packets subjected
to interaction noise. The values of breathing period are collected
from numerical simulations, which have a perfect agreement with equation
(\ref{eq:Period}). \label{tab:ShapeBreathingPeriod} }
\end{table*}

\begin{figure}
\begin{centering}
\includegraphics{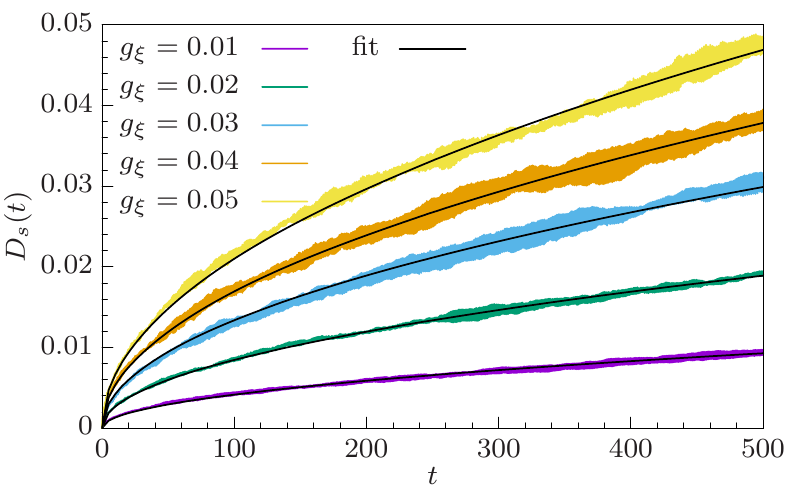}
\par\end{centering}
\caption{Shape difference evolution for non-spreading wave packet subjected
to residual interaction noise with different strength. Mean values
of shape difference $D_{s}\left(t\right)$ for 500 individual simulations
are plotted for interaction noises with strength $g_{\xi}=0.01,0.02,0.03,0.04,0.05$
(represented by different colors as labeled in the figure). The black
lines are corresponding square root function $D_{s}\left(t\right)=D_{\xi}\sqrt{t}$
fits of the data, the fitting parameters are $D_{\xi}=4.23,8.45,13.4,16.9,21.2\times10^{-4}$
respectively. The wave packet parameters are $m=1$, $l=1$ for all
lines.\label{fig:Differentg}}
\end{figure}

\begin{figure}
\begin{centering}
\includegraphics{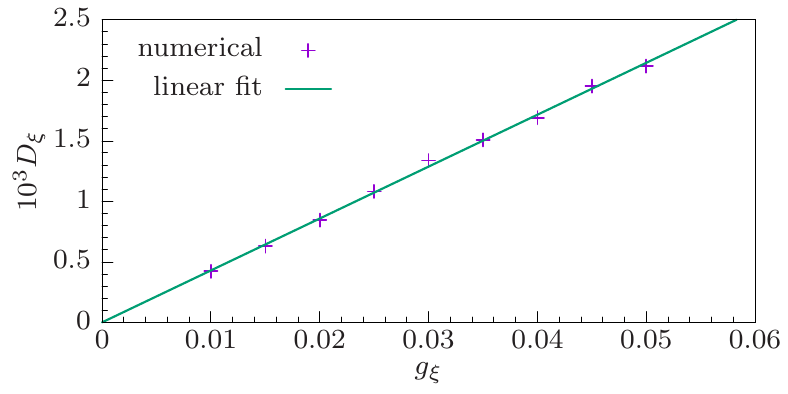}
\par\end{centering}
\caption{Shape-keeping ability of non-spreading wave packets against residual interaction
noise strength. The ``+'' are data points of $\left(g_{\xi},D_{\xi}\right)$
obtained from numerical results. The solid line is a linear fit of
the data points. The wave packet parameters are $m=1$ and $l=1$.
\label{fig:Dxigxi}}
\end{figure}

To quantitatively measure the shape stability of the non-spreading
wave packets, we introduce the following quantity
\begin{equation}
D_{s}\left(t\right)=\frac{\int\left|\left|\psi\left(\theta-\theta_{c,t},t\right)\right|^{2}
-\left|\psi\left(\theta,0\right)\right|^{2}\right|d\theta}
{\int\left|\psi\left(\theta,0\right)\right|^{2}d\theta},\label{eq:ShapeDiff}
\end{equation}
to describe the shape difference of wave packet between time $t$
and $0$ (initial). Here wave packet at time $t$ is shifted by
$\theta_{c,t}$ (azimuthal angle the wave packet have passed since
$t=0$) to line up with the initial wave packet. If $D_{s}=0$, profiles
of the wave packet at time $t$ and $0$ are absolutely the same.
And a larger value of $D_{s}$ indicates a bigger shape change. It
is reasonable to expect that the amount of shape difference will
positively relate to the strength of the interaction noise, see figure
\ref{fig:Differentg} where the shape differences of non-spreading
wave packet with $m=1$, $l=1$ are plotted for different strength
of residual interaction noises. From the figure, the numerical results
also suggest a square root formula of the shape difference
\begin{equation}
D_{s}\left(t\right)=D_{\xi}\sqrt{t}.\label{eq:Ds_fit}
\end{equation}
Thus, $D_{\xi}$ can be interpreted as a parameter to describe the shape-keeping ability
 of the non-spreading wave packets. A larger value of $D_{\xi}$
indicates that wave packet will deviate from its initial shape more
quickly, while a smaller value of $D_{\xi}$ means the wave packet
will keep its initial shape for a longer time. Numerical results show
a linear dependence of $D_{\xi}$ on the interaction noise strength
$g_{\xi}$, see figure \ref{fig:Dxigxi}. More numerical results show
that these conclusions also hold for non-spreading wave packets with other
values of $m$ and $l$.

\begin{figure}
\begin{centering}
\includegraphics{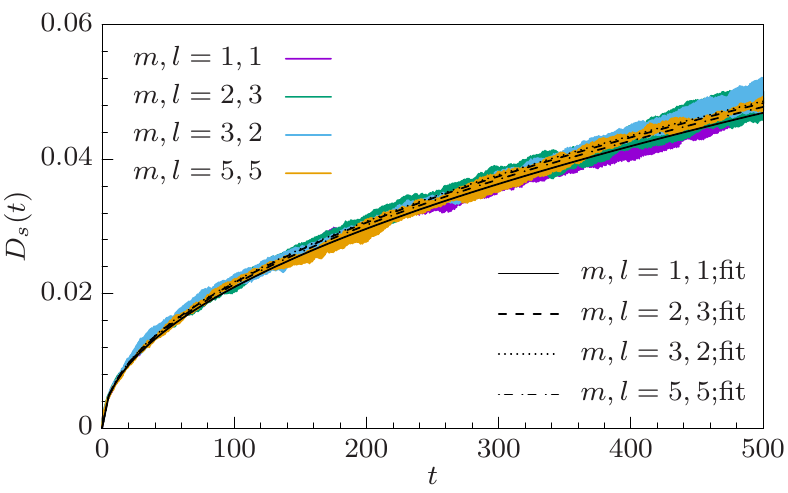}
\par\end{centering}
\caption{Shape difference evolution for different non-spreading states. Mean values of
shape difference for 500 individual simulations are plotted for different
states $m,l=1,1;2,3;3,2;5,5$ (represented by different colors as labeled
in the figure). The black lines are the corresponding square root function
$D_{s}\left(t\right)=D_{\xi}\sqrt{t}$ fits of the data, the fitting
parameters are $D_{\xi}=2.09,2.16,2.18,2.14\times10^{-3}$ respectively.
The interaction noise strength is $g_{\xi}=0.05$ for all lines.\label{fig:differentml}}
\end{figure}

We also examined the shape differences for different non-spreading
wave packets during their evolution. In figure \ref{fig:differentml},
we plot the evolution of $D_{s}$ for wave packets with parameters
$m,l=1,1;2,3;3,2;5,5$. As all the lines almost overlap with each
other and at the same time the square root function fitting parameter
$D_{\xi}=2.09,2.16,1.18,2.14\times10^{-3}$ are very close to each
other, we conclude that all the non-spreading wave packets are equally
stable against residual interaction noise.

\begin{figure}
\begin{centering}
\includegraphics{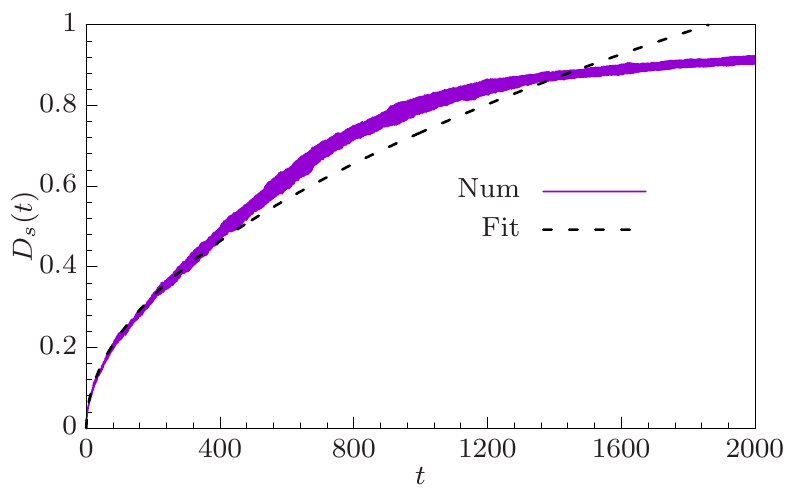}
\par\end{centering}
\caption{Saturation of shape difference. Mean values of shape difference
$D_s$ for 500 individual simulations are plotted for wave packets
$m=1,l=1$. The strength of residual interaction noise is $g_{\xi}=0.5$.
The solid line is the numerical results, and the black dash line
is the square root fit. \label{fig:saturation}}
\end{figure}

At last, we point out that the square root increasing formula (\ref{eq:Ds_fit})
only holds for a small value of $D_s$. From the definition, the value of $D_s$
will never be larger than 1. Therefore, when $D_s$ becomes large, it will
attain a saturation. In figure \ref{fig:saturation}, an example is shown
for parameters $g_{\xi}=0.5, m=1, l=1$. From the figure, one see that
when $t<400$ formula (\ref{eq:Ds_fit}) fits the data well. However,
when $t>1400$ the increase of $D_s$ becomes saturated.

\section{Conclusion\label{sec:Conclusion}}

In conclusion, we examined the non-spreading wave packets in a ring.
We found that there only exist one set of non-spreading wave packets
in a ring, and they have been found analytically. The realization
of these packets in ultra-cold atom systems (with the assistance of
Feshbach resonance technique to eliminate contact interaction between
atoms) is discussed. The stability of these wave packets against
residual interaction noise is examined numerically. It is found that
these non-spreading wave packets are stable against weak interaction
noise, and a strong interaction noise will induce a periodical shape
breathing of the wave packets. All these wave packets have the shape-keeping
ability. And the shape-keeping ability is linearly related to
the interaction noise strength.

While we focus on the non-spreading wave packet and its main features in
the present work, it will also be interesting to further build
a matter-wave interferometer using such non-spreading wave packets.
Due to the linearity feature, the splitting and recombining processes
in operating an interferometer are expected to not destroy the
non-spreading property. The detailed splitting, recombining scheme,
and the interference pattern may be discussed in future work.
Other interesting extensions of the present work are to study
the persistent current and quantum time crystal related phenomena.

\section*{Acknowledgments}

This work is supported by National Natural Science Foundation of China
(Grant No. 11847059 and 11874127).


\section*{References}
\begin{thebibliography}{10}

\bibitem{Gangat2005} Gangat A, Pradhan P, Pati G and Shahriar M S 2005 Two-dimensional nanolithography using atom interferometry \textit{Phys. Rev. A} \href{https://doi.org/10.1103/PhysRevA.71.043606}{\textbf{71} 043606}

\bibitem{Fouda2016} Fouda M F, Fang R, Ketterson J B and Shahriar M S 2016 Generation of arbitrary lithographic patterns using Bose-Einstein-condensate interferometry \textit{Phys. Rev. A} \href{https://doi.org/10.1103/PhysRevA.94.063644}{\textbf{94} 063644}

\bibitem{Dwyer2005} O\textquoteright Dwyer C, Gay G, de Lesegno B V, Weiner J, Camposeo A, Tantussi F, Fuso F, Allegrini M and Arimondo E 2005 Atomic nanolithography patterning of submicron features: writing an organic self-assembled monolayer with cold, bright Cs atom beams \textit{Nanotechnology} \href{https://doi.org/10.1088/0957-4484/16/9/022}{\textbf{16} 1536}

\bibitem{Vengalattore2007} Vengalattore M, Higbie J M, Leslie S R, Guzman J, Sadler L E and Stamper-Kurn D M 2007 High-resolution magnetometry with a spinor Bose-Einstein condensate \textit{Phys. Rev. Lett.} \href{https://doi.org/10.1103/PhysRevLett.98.200801}{\textbf{98} 200801}

\bibitem{Muessel2014} Muessel W, Strobel H, Linnemann D, Hume D B and Oberthaler M K 2014 Scalable spin squeezing for quantum-enhanced magnetometry with Bose-Einstein condensates \textit{Phys. Rev. Lett.} \href{https://doi.org/10.1103/PhysRevLett.113.103004}{\textbf{113} 103004}

\bibitem{Cronin2009} Cronin A D and Schmiedmayer J 2009 Optics and interferometry with atoms and molecules \textit{Rev. Mod. Phys.} \href{http://dx.doi.org/10.1103/RevModPhys.81.1051}{\textbf{81} 1051}

\bibitem{Lee2012} Lee C, Huang J, Deng H, Dai H and Xu J 2012 Nonlinear quantum interferometry with Bose condensed atoms \textit{Frontiers of Physics} \href{https://doi.org/10.1007/s11467-011-0228-6}{\textbf{7} 109}

\bibitem{Muntinga2013} M\"{u}ntinga H, Ahlers H, Krutzik M, Wenzlawski A, Arnold S, Becker D, Bongs K, Dittus H, Duncker H, Gaaloul N, Gherasim C, Giese E, Grzeschik C, H\"{a}nsch T W, Hellmig O, Herr W, Herrmann S, Kajari E, Kleinert S, L\"{a}mmerzahl C, Lewoczko-Adamczyk W, Malcolm J, Meyer N, Nolte R, Peters A, Popp M, Reichel J, Roura A, Rudolph J, Schiemangk M, Schneider M, Seidel S T, Sengstock K, Tamma V, Valenzuela T, Vogel A, Walser R, Wendrich T, Windpassinger P, Zeller W, van Zoest T, Ertmer W, Schleich W P and Rasel E M 2013 Interferometry with Bose-Einstein condensates in microgravity \textit{Phys. Rev. Lett.} \href{https://doi.org/10.1103/PhysRevLett.110.093602}{\textbf{110} 093602}

\bibitem{Robins2013} Robins N P, Altin P A, Debs J E and Close J D 2013 Atom lasers: production, properties and prospects for precision inertial measurement \textit{Physics Reports} \href{https://doi.org/10.1016/j.physrep.2013.03.006}{\textbf{529} 265}

\bibitem{Bolpasi2014} Bolpasi V, Efremidis N K, Morrissey M J, Condylis P C, Sahagun D, Baker M and von Klitzing W 2014 An ultra-bright atom laser \textit{New Journal of Physics} \href{http://dx.doi.org/10.1088/1367-2630/16/3/033036}{\textbf{16} 033036}

\bibitem{Khaykovich2002} Khaykovich L, Schreck F, Ferrari G, Bourdel T, Cubizolles J, Carr L D, Castin Y and Salomon C 2002 Formation of a matter-wave bright soliton \textit{Science} \href{https://doi.org/10.1126/science.1071021}{\textbf{296} 1290}

\bibitem{Strecker2002} Strecker K E, Partridge G B, Truscott A G and Hulet R G 2002 Formation and propagation of matter-wave soliton trains \textit{Nature} \href{https://doi.org/10.1038/nature747}{\textbf{417} 150}

\bibitem{Strecker2003} Strecker K E, Partridge G B, Truscott A G and Hulet R G 2003 Bright matter wave solitons in Bose-Einstein condensates \textit{New Journal of Physics} \href{https://doi.org/10.1088/1367-2630/5/1/373 }{\textbf{5} 73}

\bibitem{Abdullaev2008} Abdullaev F K and Garnier J 2008 Bright solitons in Bose-Einstein condensates: theory \textit{Emergent Nonlinear Phenomena in Bose-Einstein Condensates: Theory and Experiment} ed P G Kevrekidis, D J Frantzeskakis and R Carretero-Gonzalez (Berlin, Heidelberg: Springer Berlin Heidelberg) \href{https://doi.org/10.1007/978-3-540-73591-5_2}{p 25}

\bibitem{Martin2012} Martin A D and Ruostekoski J 2012 Quantum dynamics of atomic bright solitons under splitting and recollision, and implications for interferometry \textit{New Journal of Physics} \href{https://doi.org/10.1088/1367-2630/14/4/043040}{\textbf{14} 043040}

\bibitem{Polo2013} Polo J and Ahufinger V 2013 Soliton-based matter-wave interferometer \textit{Phys. Rev. A} \href{https://doi.org/10.1103/PhysRevA.88.053628}{\textbf{88} 053628}

\bibitem{Gertjerenken2013} Gertjerenken B 2013 Bright-soliton quantum superpositions: Signatures of high- and low-fidelity states \textit{Phys. Rev. A} \href{https://doi.org/10.1103/PhysRevA.88.053623}{\textbf{88} 053623}

\bibitem{Helm2014} Helm J L, Rooney S J, Weiss C and Gardiner S A 2014 Splitting bright matter-wave solitons on narrow potential barriers: Quantum to classical transition and applications to interferometry \textit{Phys. Rev. A} \href{https://doi.org/10.1103/PhysRevA.89.033610}{\textbf{89} 033610}

\bibitem{Sakaguchi2016} Sakaguchi H and Malomed B A 2016 Matter-wave soliton interferometer based on a nonlinear splitter \textit{New Journal of Physics} \href{https://doi.org/10.1088/1367-2630/18/2/025020/meta}{\textbf{18} 025020}

\bibitem{McDonald2014} McDonald G D, Kuhn C C N, Hardman K S, Bennetts S, Everitt P J, Altin P A, Debs J E, Close J D and Robins N P 2014 Bright Solitonic Matter-Wave Interferometer \textit{Phys. Rev. Lett.} \href{https://doi.org/10.1103/PhysRevLett.113.013002}{\textbf{113} 013002}

\bibitem{McDonald2017} McDonald G D, Kuhn C C N, Hardman K S, Bennetts S, Everitt P J, Altin P A, Debs J E, Close J D and Robins N P 2017 Erratum: Bright Solitonic Matter-Wave Interferometer [Phys. Rev. Lett. 113, 013002 (2014)] \textit{Phys. Rev. Lett.} \href{https://doi.org/10.1103/PhysRevLett.118.219903}{\textbf{118} 219903}

\bibitem{Fattori2008} Fattori M, D\textquoteright Errico C, Roati G, Zaccanti M, Jona-Lasinio M, Modugno M, Inguscio M and Modugno G 2008 Atom interferometry with a weakly interacting Bose-Einstein condensate \textit{Phys. Rev. Lett.} \href{https://doi.org/10.1103/PhysRevLett.100.080405}{\textbf{100} 080405}

\bibitem{Gustavsson2008} Gustavsson M, Haller E, Mark M J, Danzl J G, Rojas-Kopeinig G and N\"{a}gerl H-C 2008 Control of interaction-induced dephasing of Bloch oscillations \textit{Phys. Rev. Lett.} \href{https://doi.org/10.1103/PhysRevLett.100.080404}{\textbf{100} 080404}

\bibitem{Haine2016} Haine S A 2016 Mean-Field Dynamics and Fisher Information in Matter Wave Interferometry \textit{Phys. Rev. Lett.} \href{https://doi.org/10.1103/PhysRevLett.116.230404}{\textbf{116} 230404}

\bibitem{Haine2018} Haine S A 2018 Quantum noise in bright soliton matterwave interferometry \textit{New Journal of Physics} \href{https://doi.org/10.1088/1367-2630/aab47f}{\textbf{20} 33009}

\bibitem{Jo2007} Jo G-B, Shin Y, Will S, Pasquini T A, Saba M, Ketterle W, Pritchard D E, Vengalattore M and Prentiss M 2007 Long Phase Coherence Time and Number Squeezing of Two Bose-Einstein Condensates on an Atom Chip \textit{Phys. Rev. Lett.} \href{https://doi.org/10.1103/PhysRevLett.98.030407}{\textbf{98} 030407}

\bibitem{Berrada2013}  Berrada T, van Frank S, B\"{u}cker R, Schumm T, Schaff J-F and Schmiedmayer J 2013 Integrated Mach-Zehnder interferometer for Bose-Einstein condensates \textit{Nature Communications} \href{https://doi.org/10.1038/ncomms3077}{\textbf{4} 2077}

\bibitem{Gross2010} Gross C, Zibold T, Nicklas E, Est\'{e}ve J and Oberthaler M K 2010 Nonlinear atom interferometer surpasses classical precision limit \textit{Nature} \href{https://doi.org/10.1038/nature08919}{\textbf{464} 1165}

\bibitem{Lucke2011} L\"{u}cke B, Scherer M, Kruse J, Pezz\'{e} L, Deuretzbacher F, Hyllus P, Topic O, Peise J, Ertmer W, Arlt J, Santos L, Smerzi A and Klempt C 2011 Twin Matter Waves for Interferometry Beyond the Classical Limit \textit{Science} \href{https://doi.org/10.1126/science.1208798}{\textbf{334} 773}

\bibitem{Berry1979} Berry M V and Balazs N L 1979 Nonspreading wave packets \textit{American Journal of Physics} \href{https://doi.org/10.1119/1.11855}{\textbf{47} 264}

\bibitem{Hu2012} Hu Y, Siviloglou G A, Zhang P, Efremidis N K, Christodoulides D N and Chen Z 2012 Self-accelerating Airy beams: generation, control, and applications \textit{Nonlinear Photonics and Novel Optical Phenomena} (Springer Series in Optical Sciences vol 170) ed Z Chen and R Morandotti (New York, NY: Springer New York) \href{https://doi.org/10.1007/978-1-4614-3538-9_1}{p 1}

\bibitem{Siviloglou2007} Siviloglou G A, Broky J, Dogariu A and Christodoulides D N 2007 Observation of accelerating Airy beams \textit{Phys. Rev. Lett.} \href{https://doi.org/10.1103/PhysRevLett.99.213901}{\textbf{99} 213901}

\bibitem{Voloch-Bloch2013} Voloch-Bloch N, Lereah Y, Lilach Y, Gover A and Arie A 2013 Generation of electron Airy beams \textit{Nature} \href{https://doi.org/10.1038/nature11840}{\textbf{494} 331}

\bibitem{Efremidis2013} Efremidis N K, Paltoglou V and von Klitzing W 2013 Accelerating and abruptly autofocusing matter waves \textit{Phys. Rev. A} \href{https://doi.org/10.1103/PhysRevA.87.043637}{\textbf{87} 043637}

\bibitem{Yuce2015} Yuce C 2015 Self-accelerating matter waves \textit{Modern Physics Letters B} \href{https://doi.org/10.1142/S0217984915501717}{\textbf{29} 1550171}

\bibitem{Bouyer2014} Bouyer P 2014 The centenary of Sagnac effect and its applications: from electromagnetic to matter waves \textit{Gyroscopy and Navigation} \href{https://doi.org/10.1134/S2075108714010039}{\textbf{5} 20}

\bibitem{Barrett2014} Barrett B, Geiger R, Dutta I, Meunier M, Canuel B, Gauguet A, Bouyer P and Landragin A 2014 The Sagnac effect: 20 years of development in matter-wave interferometry \textit{Comptes Rendus Physique} \href{https://doi.org/10.1016/j.crhy.2014.10.009}{\textbf{15} 875}

\bibitem{Dayon2010} Dayon D J, Toland J R E and Search C P 2010 Atom gyroscope with disordered arrays of quantum rings \textit{Journal of Physics B: Atomic, Molecular and Optical Physics} \href{https://doi.org/10.1103/10.1088/0953-4075/43/11/115302}{\textbf{43} 115302}

\bibitem{Helm2015} Helm J L, Cornish S L and Gardiner S A 2015 Sagnac interferometry using bright matter-wave solitons \textit{Phys. Rev. Lett.} \href{https://doi.org/10.1103/PhysRevLett.114.134101}{\textbf{114} 134101}

\bibitem{Stevenson2015} Stevenson R, Hush M R, Bishop T, Lesanovsky I and Fernholz T 2015 Sagnac interferometry with a single atomic clock \textit{Phys. Rev. Lett.} \href{https://doi.org/10.1103/PhysRevLett.115.163001}{\textbf{115} 163001}

\bibitem{Ryu2013} Ryu C, Blackburn P W, Blinova A A and Boshier M G 2013 Experimental realization of Josephson junctions for an atom SQUID \textit{Phys. Rev. Lett.} \href{https://doi.org/10.1103/PhysRevLett.111.205301}{\textbf{111} 205301}

\bibitem{Wang2015} Wang Y-H, Kumar A, Jendrzejewski F, Wilson R M, Edwards M, Eckel S, Campbell G K and Clark C W 2015 Resonant wavepackets and shock waves in an atomtronic SQUID \textit{New Journal of Physics} \href{https://doi.org/10.1088/1367-2630/17/12/125012 }{\textbf{17} 125012}

\bibitem{Mathey2016} Mathey A C and Mathey L 2016 Realizing and optimizing an atomtronic SQUID \textit{New Journal of Physics} \href{https://doi.org/10.1088/1367-2630/18/5/055016 }{\textbf{18} 055016}

\bibitem{Kumar2016} Kumar A, Anderson N, Phillips W D, Eckel S, Campbell G K and Stringari S 2016 Minimally destructive, Doppler measurement of a quantized flow in a ring-shaped Bose-Einstein condensate \textit{New Journal of Physics} \href{https://doi.org/10.1088/1367-2630/18/2/025001}{\textbf{18} 025001}

\bibitem{Beattie2013} Beattie S, Moulder S, Fletcher R J and Hadzibabic Z 2013 Persistent currents in spinor condensates \textit{Phys. Rev. Lett.} \href{https://doi.org/10.1103/PhysRevLett.110.025301}{\textbf{110} 025301}

\bibitem{Ryu2007} Ryu C, Andersen M F, Clad\'{e} P, Natarajan V, Helmerson K and Phillips W D 2007 Observation of persistent flow of a Bose-Einstein condensate in a toroidal trap \textit{Phys. Rev. Lett.} \href{https://doi.org/10.1103/PhysRevLett.99.260401}{\textbf{99} 260401}

\bibitem{Brand2001} Brand J and Reinhardt W P 2001 Generating ring currents, solitons and svortices by stirring a Bose-Einstein condensate in a toroidal trap \textit{Journal of Physics B: Atomic, Molecular and Optical Physics} \href{https://doi.org/10.1088/0953-4075/34/4/105}{\textbf{34} L113}

\bibitem{Li2012} Li T, Gong Z-X, Yin Z-Q, Quan H T, Yin X, Zhang P, Duan L-M and Zhang X 2012 Space-time crystals of trapped ions \textit{Phys. Rev. Lett.} \href{https://doi.org/10.1103/PhysRevLett.109.163001}{\textbf{109} 163001}

\bibitem{Ohberg2018} \"{O}hberg P and Wright E M 2018 On quantum time crystals and interacting gauge theories in atomic Bose-Einstein condensates \href{https://arxiv.org/abs/1812.04672}{arXiv:1812.04672}

\bibitem{Sacha2018} Sacha K and Zakrzewski J 2018 Time crystals: a review \textit{Reports on Progress in Physics} \href{https://doi.org/10.1088/1361-6633/aa8b38}{\textbf{81} 16401}

\bibitem{Chin2010} Chin C, Grimm R, Julienne P and Tiesinga E 2010 Feshbach resonances in ultracold gases \textit{Rev. Mod. Phys.} \href{https://doi.org/10.1103/RevModPhys.82.1225}{\textbf{82} 1225}

\bibitem{Timmermans1999} Timmermans E, Tommasini P, Hussein M and Kerman A 1999 Feshbach resonances in atomic Bose-Einstein condensates \textit{Physics Reports} \href{https://doi.org/10.1016/S0370-1573(99)00025-3}{\textbf{315} 199}

\bibitem{Pollack2009} Pollack S E, Dries D, Junker M, Chen Y P, Corcovilos T A and Hulet R G 2009 Extreme Tunability of Interactions in a $^7$Li Bose-Einstein Condensate \textit{Phys. Rev. Lett.} \href{https://doi.org/10.1103/PhysRevLett.102.090402}{\textbf{102} 090402}

\bibitem{Madelung1927} Madelung E 1927 Quantentheorie in hydrodynamischer Form \textit{Zeitschrift f\"{u}r Physik} \href{https://doi.org/10.1007/BF01400372}{\textbf{40} 322}

\bibitem{Lin2008} Lin C-L, Huang M-J and Hsiung T-C 2008 Nonspreading wave packets in a general potential V(x,t) in one dimension \href{https://arxiv.org/abs/0809.4105}{arXiv:0809.4105}

\bibitem{Wright2000} Wright E M, Arlt J and Dholakia K 2000 Toroidal optical dipole traps for atomic Bose-Einstein condensates using Laguerre-Gaussian beams \textit{Phys. Rev. A} \href{https://doi.org/10.1103/PhysRevA.63.013608}{\textbf{63} 13608}

\bibitem{Arnold2004} Arnold A S 2004 Adaptable-radius, time-orbiting magnetic ring trap for Bose-Einstein condensates \textit{Journal of Physics B: Atomic, Molecular and Optical Physics} \href{https://doi.org10.1088/0953-4075/37/2/l03 }{\textbf{37} L29}

\bibitem{Crookston2005} Crookston M B, Baker P M  and Robinson M P 2005 A microchip ring trap for cold atoms \textit{Journal of Physics B: Atomic, Molecular and Optical Physics} \href{https://doi.org/10.1088/0953-4075/38/18/001}{\textbf{38} 3289}

\bibitem{Heathcote2008} W H Heathcote W H, Nugent E,Sheard B T and Foot C J 2008 A ring trap for ultracold atoms in an RF-dressed state \textit{New Journal of Physics} \href{https://doi.org/10.1088/1367-2630/10/4/043012} {\textbf{10}, 043012}

\bibitem{Gardiner1985} Gardiner C W 1985 \textit{Handbook of stochastic methods for physics, chemistry, and the natural sciences} (Berlin, Heidelberg, New York: Springer-Verlag)
\end{thebibliography}
\end{document}